\documentclass[12pt]{article}
\usepackage{graphicx}

\def \bea{\begin{eqnarray}}
\def \beq{\begin{equation}}

\def \eea{\end{eqnarray}}
\def \eeq{\end{equation}}

\begin{document}
\rightline{CLNS 03/1856}
\rightline{hep-ph/0312269}
\bigskip

\Large
\centerline{\bf Exotic states of matter in heavy meson decays
\footnote{Submitted to Phys.\ Rev.\ D.}}
\normalsize
\bigskip
 
\centerline{Jonathan L. Rosner~\footnote{rosner@hep.uchicago.edu.  On leave
from Enrico Fermi Institute and Department of Physics,
University of Chicago, 5640 S. Ellis Avenue, Chicago, IL 60637}}
\centerline{Laboratory of Elementary Particle Physics}
\centerline{Cornell University, Ithaca, NY 14850}
 
\begin{quote}

The potential of decays of mesons containing heavy quarks [including $B$
mesons and the $\Upsilon(1S)$] for producing final states of matter with
unusual quark configurations, such as $q q \bar q \bar q$ or $q q q q \bar q$,
is investigated.  The usefulness of antineutron detection in such searches
is stressed.

\end{quote}

\noindent
PACS Categories: 13.25.Gv, 13.25.Hw, 14.40.Gx, 14.40.Nd

\bigskip

\centerline{\bf I.  INTRODUCTION}
\bigskip

The decays of mesons containing heavy quarks provide carefully controlled
environments for production of hadronic states with specific quantum numbers.
Copious samples of such mesons accumulated at the BaBar, Belle, and CLEO
detectors permit searches for interesting forms of matter, including mesons
beyond the usual quark-antiquark and baryons beyond the usual three-quark
configurations.  Such hadrons (e.g., $q q \bar q \bar q$ mesons and
$q q q q \bar q$ baryons) are known as {\it exotic}.  Their existence is
not forbidden by quantum chromodynamics (QCD) as long as they maintain
color-singlet configurations.  They are predicted in a wide variety of schemes.

In the present paper, which follows up some suggestions made in Ref.\
\cite{Rosner:2003bm} and presents others, we explore some ways in which the
decays of mesons containing heavy (mainly $b$) quarks can help in the search
for exotic hadrons.  We focus on states which are manifestly exotic by virtue
of their internal quantum numbers, bypassing the question of whether certain
recently seen resonances are heavy quarkonium states or molecules of flavored
mesons.  Our discussion of decays of particles containing charmed ($c$) quarks
will be brief since they are much lighter, thereby probing a much
more limited mass range.

We review some of the theoretical framework for exotic states in Section II.
We then discuss exotic meson and baryon final states of $B$ decays in
Sections III and IV, respectively.  Section V is devoted to some specific
signatures of a recently discussed candidate for a light $uudd \bar s$
baryon.  Possibilities for exotic particle observation in $\Upsilon(1S)$
decays are noted in Section VI.  Some remarks on charmed particle decays
are made in Section VII, while Section VIII summarizes.
\bigskip

\centerline{\bf II.  THEORETICAL SCHEMES}
\bigskip

The duality of scattering amplitudes in crossed channels \cite{Freund:1967hw}
was used \cite{Rosner:1968si} to argue in favor of the existence of $q q \bar q
\bar q$ mesons coupling mainly to baryon-antibaryon final states.  (For a
recent review of this work, see Ref.\ \cite{Roy:2003hk}.)  These arguments are
conveniently visualized using quark graphs \cite{Harari:nn}.  Generalizations
of the graphical arguments \cite{Roy:sb} lead to selection
rules governing allowed transitions among exotic states.

The properties of $q q \bar q \bar q$ mesons were analyzed within a particular
confinement scheme, known as the MIT Bag Model \cite{Chodos:1974je}, by Jaffe
\cite{Jaffe:1976ig}.  Masses and couplings to meson pairs were calculated;
there was no particular reason for such states to favor coupling to
baryon-antibaryon channels.

A complementary approach investigated the
properties of baryon-antibaryon bound states from a standpoint primarily
motivated by nuclear physics arguments; see Ref.\ \cite{Richard:1999qh} for
a brief historical review.  Attempts to describe ordinary mesons such as
the pion in terms of such bound states have a long and eminent history
\cite{Fermi:sa}.  Recently a candidate for such a state
near nucleon-antinucleon threshold has been reported in $J/\psi$ decays
\cite{Bai:2003sw}.  Ref.\ \cite{Rosner:2003bm} discusses possible
interpretations of this state as well as some earlier candidates for
nucleon-antinucleon resonances.

Bootstrap-like models of mesons and baryons, which predict properties of
excited states by taking account of mesonic rather than quark degrees of
freedom, generally do not limit the quantum numbers of states to those
of $q \bar q$ or $qqq$.  An early example is the strong-coupling theory
\cite{Goebel}, generalizing Chew and Low's theory of the $\Delta^{++}$ isobar
\cite{Chew:1955zz}.  States with $I = J = 1/2,~3/2,~5/2. \ldots$ are predicted
in this approach.  More recently, chiral soliton models of baryons
\cite{Skyrme} have taken account of mesonic rather than quark degrees of
freedom.  In flavor SU(3), several authors noted that these approaches led to
an antidecuplet with $J^P = 1/2^+$ lying not far above the $1/2^+$ octet and
$3/2^+$ decuplet \cite{Manohar:1984ys,Diakonov:1997mm}.  The lightest member
of this ${\overline{\bf 10}}$ representation was expected to have 
strangeness $S=+1$, and a mass of 1530 MeV/$c^2$ was predicted
\cite{Diakonov:1997mm,Praszalowicz:2003ik}.
As will be noted in Sec.\ V, the search for this
state has taken on new urgency as a result of several claims for its
observation.  First, however, we note some special features of heavy meson
decays that make them particularly useful in the search for exotic mesons
and baryons.
\bigskip

\centerline{\bf III.  EXOTIC MESONS IN {\boldmath $B$} DECAYS}
\bigskip

We shall be concerned in this Section with the dominant subprocess $\bar b \to
\bar c u \bar d$ in $B$ meson decays.  The fact that this leads to three quarks
of different flavor in the final state provides an immediate advantage in the
search for exotic particles.  (The favored decays of charm, $c \to s u \bar d$,
share this property.)

\begin{figure}
\begin{center}
\includegraphics[height=3.7in]{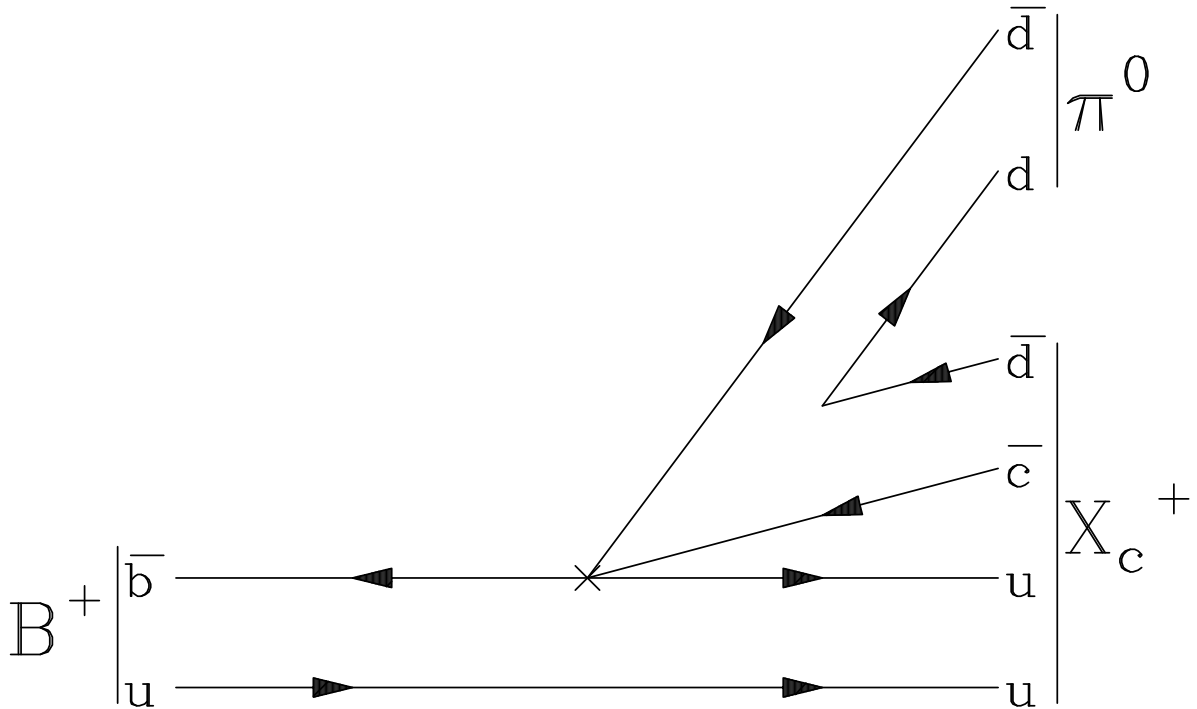}
\caption{Production of an exotic meson $X_c^{+} = uu \bar c \bar d$ in $B^+$
decays.  The $\pi^0$ could also be a photon or any other neutral particle.
Here and subsequently the weak vertex is denoted by $\times$.
\label{fig:xcpp}}
\end{center}
\end{figure}

If a $\bar b$ quark is bound to a $u$ quark in a $B^+$ meson, the favored
nonleptonic final state is then $\bar c u \bar d u$.  This, by itself, is
exotic.  Thus one means of searching for exotics is simply the study of the
missing mass of $X_c^+$ in $B^+ \to (\gamma,\pi^0,\eta,\eta',\ldots) + X_c^+$.
A quark diagram associated with this process is shown in Fig.\ \ref{fig:xcpp}.
Another diagram can be drawn in which the final $u \bar c \bar d$ decay
products of $\bar b$ are incorporated into the $X_c^+$ along with a $u$ from
$u \bar u$ production.  Similarly, the subsystem $X_c^+$ in $B^0 \to \pi^-
X_c^+$ will be exotic, also carrying the quantum numbers of $\bar c u \bar d
u$.

\begin{figure}
\begin{center}
\includegraphics[height=3.55in]{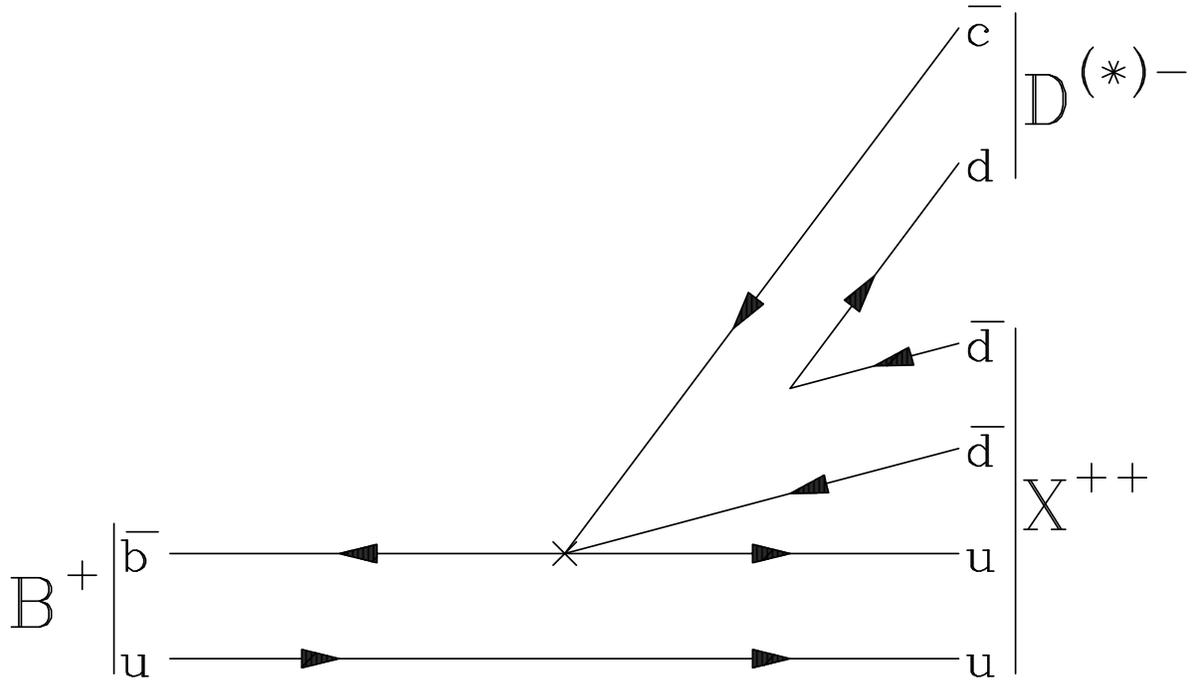}
\caption{Production of an exotic meson $X^{++} = uu \bar d \bar d$ in $B^+$
decays.  If the recoiling charmed meson is a $D^{*-}$, identification may
be easier through the decay $D^{*-} \to \pi^- \bar D^0$.
\label{fig:xpp}}
\end{center}
\end{figure}

A $B^+$ decay involving $\bar b \to u \bar c \bar d$ can produce an exotic
meson consisting entirely of light quarks ($u u \bar d \bar d$) if the $u u
\bar d$ system picks up another $\bar d$ quark in fragmentation, while the
corresponding $d$ quark accompanies the $\bar c$ to form a $D^-$ or $D^{*-}$.
This process is shown in Fig.\ \ref{fig:xpp}.  The detection of an exotic final
state in the decay $B^+ \to X^{++} D^{(*)-}$ might proceed in several ways.

(1) One could study the missing mass opposite the $D^{(*)-}$, if the energy and
momentum of the $B^+$ are known.  Such information might be provided by
reconstructing the opposite-side $B^-$ in the decay $\Upsilon(4S) \to B^+
B^-$.

(2) One could look for a mass peak in the decay products of $X^{++}$.  If
$X^{++}$ indeed decays predominantly to baryon-antibaryon final states, as
proposed in some schemes (e.g., \cite{Rosner:1968si}), one could look for
$X^{++} \to \Delta^{++} \bar \Delta^{0} \to p \pi^+ \bar p \pi^+$.

(3) One could use a combination of methods (1) and (2) if some of the decay
products of $X^{++}$ are difficult to observe, as in $X^{++} \to \Delta^{++}
\bar n$.  Here one might look for annihilation of the $\bar n$ in the
detector.  Such a signal has attracted attention as a possible background
to $b \to s \gamma$ \cite{Chen:2001fj}, and was utilized in a study of $B^0 \to
D^{*-} p \bar n$ \cite{Anderson:2000tz}. While the energy of the $\bar n$ might
not be well-measured, one could use information about the remaining particles
in the event to constrain it.

It is possible that $X^{++}$ does not decay predominantly to baryon-antibaryon
final states, for example if it lies in the below-threshold mass range
predicted in Ref.\ \cite{Jaffe:1976ig}.  One final state which has not been
thoroughly explored for exotic mesons, and which is observable by modern
detectors such as BaBar, Belle, and CLEO, is $\rho^+ \rho^+$.  An $I=2$ meson
with favored coupling to $\rho \rho$ and mass 1.4--1.6 GeV/$c^2$
could help to explain the predominance of $\gamma \gamma \to \rho^0 \rho^0$
over $\gamma \gamma \to \rho^+ \rho^-$ near threshold \cite{Achasov:1981kh}.

\begin{figure}
\begin{center}
\includegraphics[height=3.55in]{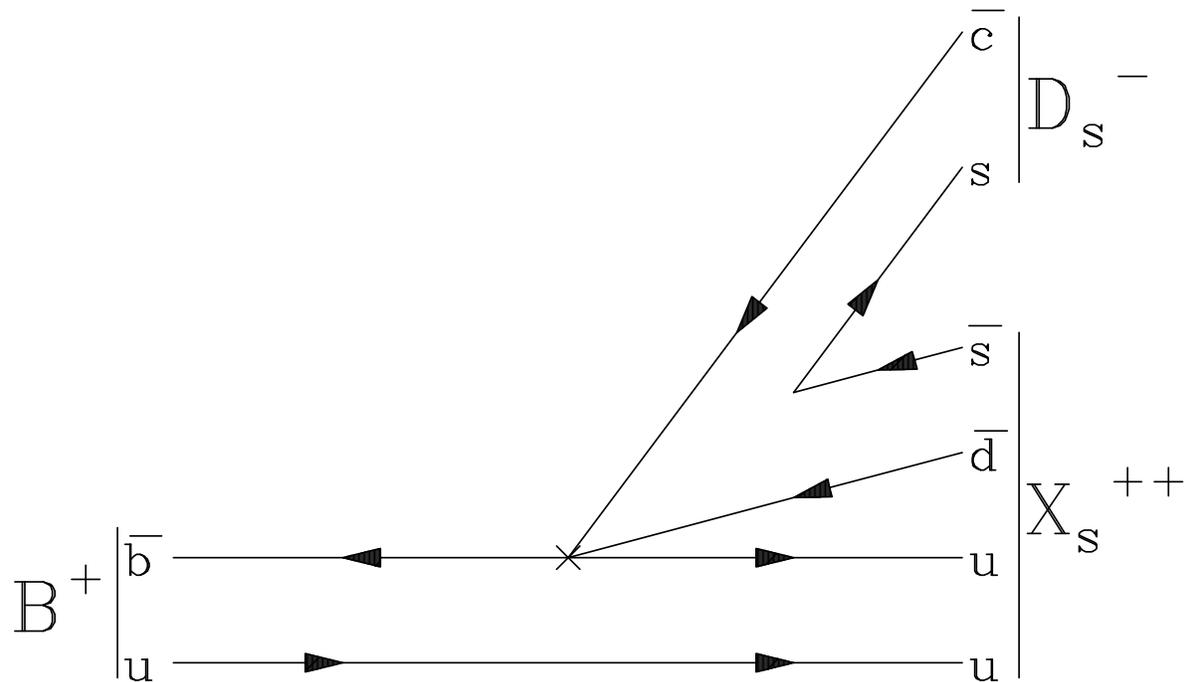}
\caption{Production of a strange exotic meson $X^{++} = uu \bar d \bar s$ in
$B^+$ decays.
\label{fig:xpps}}
\end{center}
\end{figure}

A strange exotic meson can be formed if the final $u u \bar d$ state
picks up an anti-strange $\bar s$ quark, so that the recoil system is
(for example) $D_s^-$.  This process is shown in Fig.\ \ref{fig:xpps}.

The recoil $D_s^-$ could be replaced by a $\bar K \bar D$ system.  The
$X_s^{++}$ can decay, for example, to $\bar \Lambda p \pi^+$ if
baryon-antibaryon channels are favored, or to $K^+ \pi^+ (m \pi^0)$ or
$K^+ \pi^+ (m \pi^+ m \pi^-)$ $(m=0,1,2,\ldots)$
if the lighter exotics of Ref.\ \cite{Jaffe:1976ig} are favored.

If the final $u u \bar d$ state picked up a $\bar u$ during fragmentation,
we would not be able to tell the exotic nature of the final product.  That is
why we have focused on fragmentation mechanisms involving $\bar d$ or $\bar s$.

The presence of baryon-antibaryon pairs in $B$ decays is well-documented.
Channels which have been seen include $B \to \bar D^{(*)} N \bar N$
\cite{Anderson:2000tz,Abe:2002tw}, $B^+ \to K^+ p \bar p$ \cite{Abe:2002ds},
$B^0 \to \bar \Lambda p \pi^-$ \cite{Wang:2003yi}, $B^+ \to p \bar p \pi^+,~
p \bar p K^0$, and $p \bar p K^{*+}$ \cite{Wang:2003iz}, and $B \to \Lambda_c^+
\bar p (m \pi)$ $(m=0,1,2,3)$ \cite{Dytman:2002yd,Gabyshev:2002zq}. Theoretical
investigations of such decays include \cite{Rosner:2003bm,Dunietz:1998uz,%
Hou:2000bz,Cheng:2001ub,Luo:2003pv}.  The decays being considered here may
involve somewhat higher multiplicities than those reported so far.
\bigskip

\centerline{\bf IV.  EXOTIC BARYONS IN {\boldmath $B$} DECAYS}
\bigskip

\begin{figure}
\begin{center}
\includegraphics[height=3.65in]{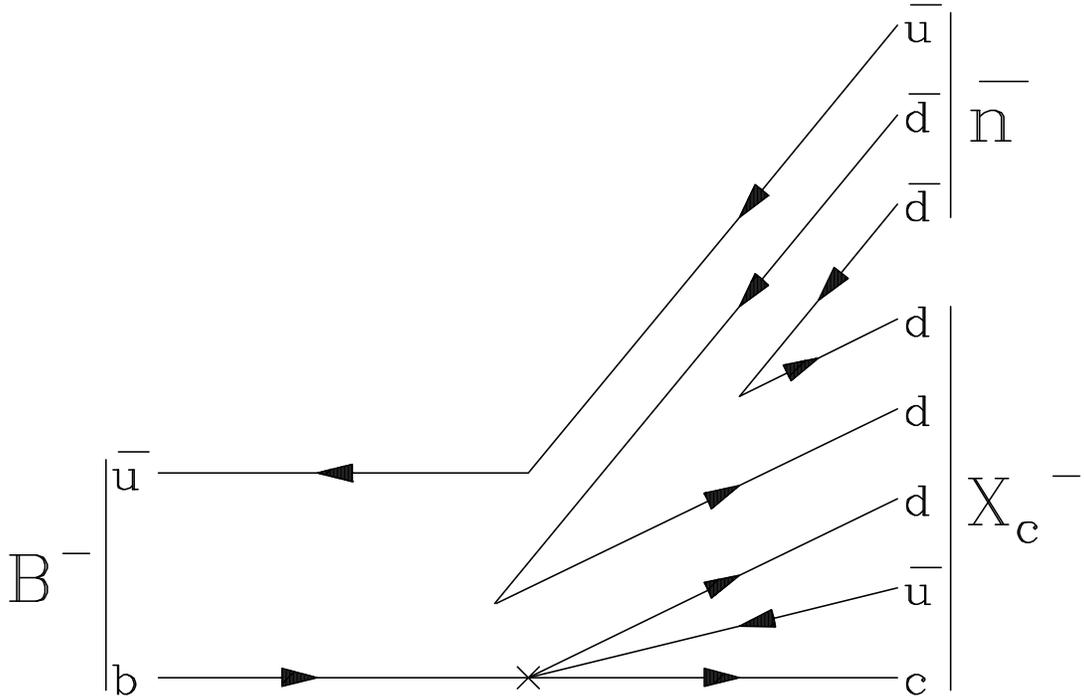}
\caption{Production of a charmed exotic baryon $X_c^{-} = cddd \bar u$ in $B^-$
decays.
\label{fig:xc}}
\end{center}
\end{figure}

\begin{figure}
\begin{center}
\includegraphics[height=3.7in]{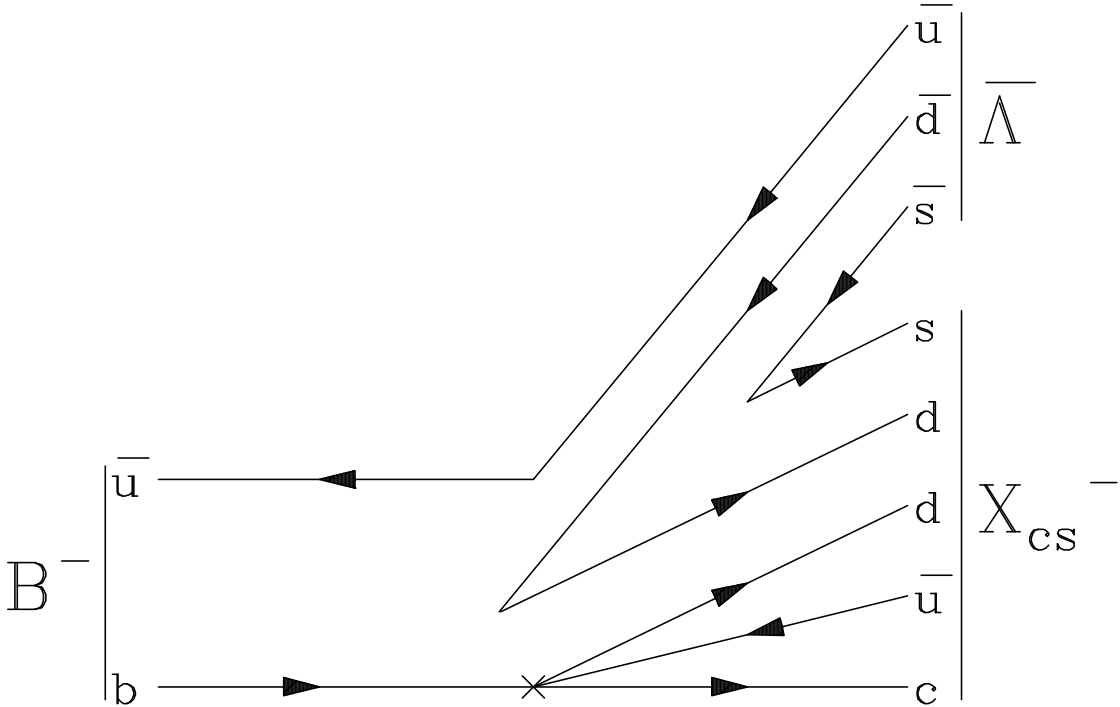}
\caption{Production of a charmed-strange exotic baryon $X_{cs}^{-} = cdds \bar
u$ in $B^-$ decays. 
\label{fig:xcs}}
\end{center}
\end{figure}

As in exotic meson production, several fragmentation mechanisms can produce
exotic baryons in $B$ decays.  One, leading to an exotic charmed
baryon, is illustrated in Fig.\ \ref{fig:xc} for a non-strange charmed baryon
$X_c^- = cddd \bar u$ and in Fig.\ \ref{fig:xcs} for a strange charmed
baryon $X_{cs}^- = cdds \bar u$.  Here the baryon contains all three quarks
produced in the decay $b \to c \bar u d$ as well as two more quarks (e.g.,
$dd$ or $ds$) picked up during fragmentation.  (If a $u$ quark were picked up
during fragmentation, we would not be able to tell that the state was exotic.)

A state composed of $cddd \bar u$ has the same quantum numbers as $\Sigma_c^0
\pi^-$ or $\Lambda_c^+ \pi^- \pi^-$.  In Fig.\ \ref{fig:xc} this state is shown
recoiling against an antineutron, the importance of whose detection we have
discussed above.  Of course, one could replace $\bar n$ by $\bar p \pi^+$,
looking at the missing mass $X^-$ in $B^- \to \bar p \pi^+ X^-$.  Indeed, the
decay $B^- \to \Lambda_c^+ \bar p \pi^+ \pi^- \pi^-$, permitting the study of
this final state, has been reported by the CLEO Collaboration
\cite{Dytman:2002yd}.

The corresponding charmed-strange state composed of $cdds \bar u$ has the same
quantum numbers as $\Xi_c^0 \pi^-$, and is shown in Fig.\ \ref{fig:xcs}
recoiling against a $\bar \Lambda$.  Thus, a useful missing-mass search
involves the decay $B^- \to \bar \Lambda X^-$.

\begin{figure}
\begin{center}
\includegraphics[height=3.7in]{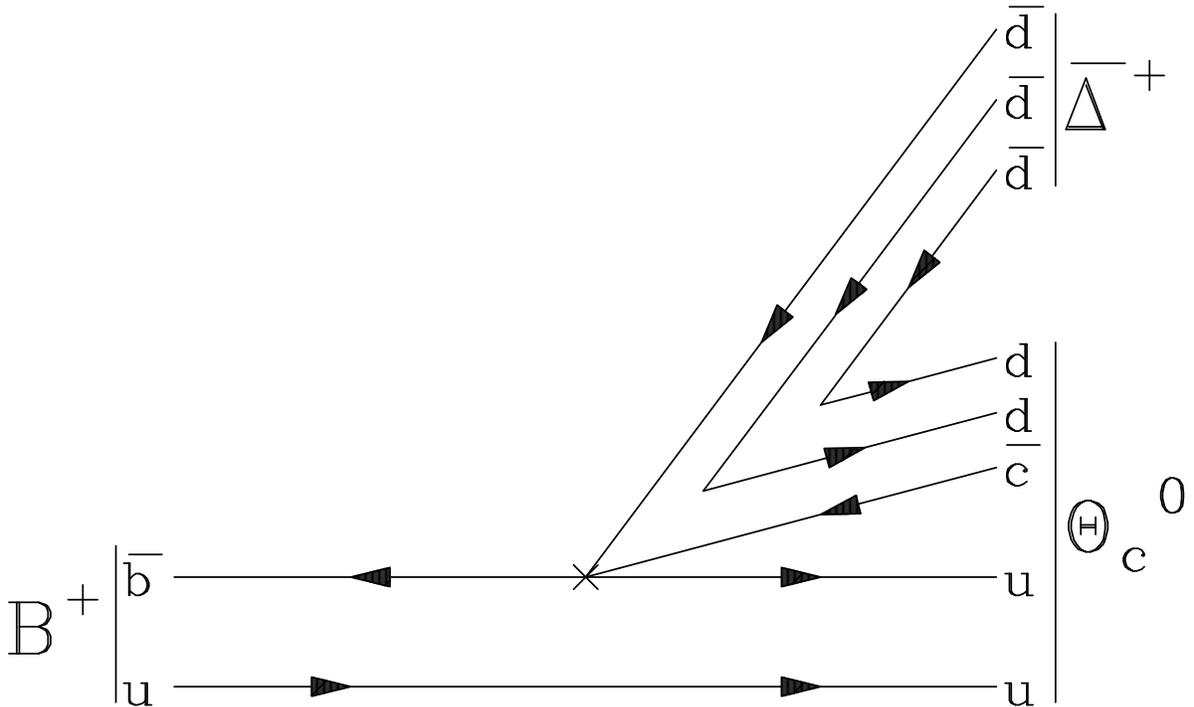}
\caption{Production of a charmed exotic baryon $\Theta_c^{0} = uudd \bar c$ in
$B^+$ decays. 
\label{fig:thc}}
\end{center}
\end{figure}

Another type of fragmentation involves a baryon containing only a subset
of the quarks in the final state of $\bar b \to \bar c u \bar d$.  An example
in which the baryon contains the $\bar c u$, a spectator $u$ from the
decaying $B^+$, and two additional $d$ quarks picked up during fragmentation,
is shown in Fig.\ \ref{fig:thc}.

In the example shown here, the exotic baryon recoils against
$\bar d \bar d \bar d = \bar \Delta^+$ which decays to $\bar n \pi^+$.  If one
$d \bar d$ pair in fragmentation is replaced by an $s \bar s$ pair, the exotic
baryon recoils against $\bar d \bar d \bar s = \bar \Sigma^+$ (or $\bar
\Lambda \pi^+$), while if both $d \bar d$ pairs are replaced by $s \bar s$, the
exotic baryon recoils against $\bar d \bar s \bar s = \bar \Xi^+$.  It may
be easier to look directly for decay products of the $\Theta_c^{0} = uudd \bar
c$, which could be $p D^-$ if the state is above that threshold.
If the state decays weakly via $\bar c \to \bar s d \bar u$, the quantum
numbers will be the same as those of $dddu \bar s$, i.e., $I_3 = -1,~S=1$.
This is still manifestly exotic.  Karliner and Lipkin \cite{Karliner:2003si}
have predicted such a state to have a mass of $2985 \pm 50$ MeV/$c^2$ and to
show up as a narrow peak in $\bar D^0 n$ and $D^- p$ spectra.  The decay
$D^{*-} p$, which has certain advantages for identification, also may be
kinematically allowed.  If the mass is light enough, a weak decay can occur
involving a detached proton vertex \cite{Lipkin:1998pb}.

Our examples have clearly not exhausted nearly all the possibilities for
exotic mesons and baryons in $B$ decays.  Those will be determined as much
by experimental opportunities as by a systematic listing of decay and recoil
modes.  Now, however, we turn our attention to a state which has been the
subject of much recent interest.
\bigskip

\centerline{\bf V. A {\boldmath $uudd \bar s$} STATE IN {\boldmath $B$} DECAYS}
\bigskip

A narrow $S=1$ baryon resonance was observed recently in photoproduction
from the neutron on $^{12}$C \cite{Nakano:2003qx} and deuterium
\cite{Stepanyan:2003qr} targets, and in photoproduction from the proton
\cite{Kubarovsky:2003fi}.  It was also reported in $K^+$ collisions
with Xe nuclei \cite{Barmin:2003vv}, in neutrino interactions as a $K_S p$
resonance \cite{Asratyan:2003cb}, and in the mass spectrum recoiling against
$\Sigma^+$ in $pp \to p K^0 \Sigma^+$ \cite{Eyrich,Hanhart:2003xp}. It has
figured prominently
in a recent conference devoted to multiquark hadrons, where further references
may be found \cite{YITP}.  The state has been dubbed
$\Theta^+$.  Its minimal quark content would be $u u d d \bar s$.

The mass of the reported state (around 1540 MeV/$c^2$) is close to that
anticipated from chiral soliton models \cite{Diakonov:1997mm,%
Praszalowicz:2003ik} for a $J^P = 1/2^+$ state.  (The prediction of Ref.\
\cite{Diakonov:1997mm} relies to some extent on identification of an SU(3)
partner of the state with an observed $1/2^+$ nucleon resonance at 1710
MeV/$c^2$.)  A $J^P = 1/2^+$ quark configuration is
proposed in Ref.\ \cite{Karliner:2003sy}.  Another (in which the state
involves two $I = 0$ $ud$ diquarks bound to an $\bar s$ quark with total $J^P =
1/2^+$) appears in \cite{Jaffe:2003sg}; see \cite{Roy:2003hk} for other
references.  The state appears to have zero isospin, showing up only in the
$K^0 p$ and $K^+ n$ final states.  Its spin and parity are not yet known
\cite{Thomas:2003ak} (see also \cite{Hanhart:2003xp}), and a lattice QCD
calculation \cite{Sasaki:2003gi} (see also \cite{Csikor:2003ng})
predicts negative parity.  The direct examination of $K^+$ charge-exchange and
total cross sections on xenon and deuterium, however, shows that if such a
resonance exists, its total width must be less than of the order of an MeV
\cite{Cahn:2003wq}, considerably less than the value of 15 MeV predicted in
Ref.\ \cite{Diakonov:1997mm}.  (See Ref.\ \cite{Praszalowicz:2003tc} for a
recent discussion of the width prediction.)  Partial-wave analyses of $K^+ N$
data \cite{Arndt:2003xz} also exclude a resonance more than a few MeV wide, and
other analyses come to similar conclusions \cite{Nussinov:2003ex}.  The
suggestion has been made that the observed effect may be a kinematic reflection
\cite{Dzierba:2003cm}.

Part of the difficulty in pinning down such a resonance comes from its
decay products.  Direct observation of a decay to $K^+ n$ requires detection of
the neutron, while direct observation of a decay to $K^0 p$ is difficult
since one only sees the $K_S$ decay mode of $K^0$.  The potential for confusion
with the non-exotic $\bar K^0 p$ channel thus exists.  (On the other hand,
there is no known evidence for a $\bar K^0 p$ resonance [a $\Sigma^*$] at 1540
MeV/$c^2$.)

Another potential source of a low-mass kinematic enhancement in the $K^+ n$
channel occurs in $K^+ K^-$ photoproduction on nuclei.  While this mechanism
does not apply to all the channels mentioned above, it is a source of concern
for some claims for a $\Theta(1540)$.

If the photoproduction of $K^+ K^-$ pairs is dominated by a virtual transition
in which the pair is kicked into a low-$M(K^+ K^-)$ state (e.g., the $\phi$) by
diffraction, the $K^+$ and $K^-$ will tend to have equal velocities $v$ with
respect to the nucleus.  Neglecting nuclear Fermi momentum, the center-of-mass
(c.m.) energy of each kaon with respect to each nucleon in the nucleus is
roughly the same.  The $K^-$ can form a prominent resonance $\Lambda(1520)$ of
spin-parity $J^P = 3/2^-$ with any proton in the nucleus.  This resonance
decays approximately 45\% of the time back to $\bar K N$ \cite{PDG}, and
is quite prominent in the $K^- p$ mass spectra in the aforementioned
photoproduction experiments.  The emerging $K^-$ from decay of this resonance
can have a quite different direction in the c.m. from the initial one, so
it no longer will form a state of low effective mass with the $K^+$.

Meanwhile the $K^+$, having approximately the same velocity in the laboratory
frame as the $K^-$, is {\it necessarily forced to have the same effective
mass in combination with any other nucleon in the nucleus} as the $K^-$
with the proton.  The argument is simplest for the deuteron, in which the
binding energy is particularly small.  Thus, when $M(K^- p) \simeq 1520$
MeV/$c^2$, one will necessarily have a low-mass peak in $M(K^+ n)$ near the
same value.  A more explicit calculation should be performed to see whether
this is indeed a problem for deuterium or other nuclei.

$B$ meson decays provide a potentially useful source of kaons and baryons
whose final-state interactions may be examined for evidence of the $\Theta^+$.
(See Ref.\ \cite{Casher:2003ep} for the suggestion that this state be
looked for in nucleon-antinucleon interactions.)  We consider charmless
decays, in which a $\bar b \to \bar s$ penguin amplitude appears to play
a large role.  Several examples illustrate the power of these decays in
searches.

Consider, for example, the process $B^0 \to p \bar p K^0$ reported in
Ref.\ \cite{Wang:2003iz}.  If one were able to flavor-tag the $B^0$ so as
to be sure it was not a $\bar B^0$, one would have an indication of the
strangeness of the neutral kaon, since $B^0$ decays generally lead to $S=1$
and not $S=-1$ states.  One could then examine the $K^0 p$
effective-mass spectrum for a low-mass peak.

No flavor tagging would be needed if one were to observe $B^0 \to K^+ n
\bar p$.  Given an observation of the neutron, one could then directly
plot the $K^+ n$ effective mass.  However, neutrons are elusive in all
detectors.

The charge-conjugate mode $\bar B^0 \to K^- \bar n p$ produces an antineutron
whose annihilation in the calorimeter of the detector may provide enough of
a constraint to permit kinematic reconstruction, especially if the
opposite-side $B^0$ is reconstructed as well.  One then would search the
$K^- \bar n$ effective-mass spectrum for evidence of the $\bar \Theta^-$.

If one can identify an antineutron in $B^+ \to K^0 p \bar n$, one can plot
the $K^0 p$ effective mass with reasonable confidence that the neutral
kaon is not a $\bar K^0$, since a $B^+$ charmless decay is much more likely to
yield an anti-strange $\bar s$ quark than an $s$ quark.  If one replaces the
$\bar n$ in this case with $\bar p \pi^+$, the reconstruction of the decay
is much more straightforward.

\begin{table}
\caption{Possible decay modes of a $\Theta^+ = uudd \bar s$ observable in
$B$ decays.  Decay products of the $\Theta^+$ are indicated in square brackets.
\label{tab:modes}}
\begin{center}
\begin{tabular}{c c c} \hline \hline
Decaying &       Mode       & Comments \\
particle &                  &          \\ \hline
$B^+$    & $[K^0 p] \bar n$       & See $\bar n$ annihilate \\
         & $[K^0 p] \bar p \pi^+$ & Full reconstruction \\
         & $[K^+ n] \bar p \pi^+$ & Only missing neutron \\
$B^-$    & $[\bar K^0 \bar p] n$  & Only missing neutron \\
         & $[\bar K^0 \bar p] p \pi^-$ & Full reconstruction \\
         & $[K^- \bar n] p \pi^-$ & See $\bar n$ annihilate \\
$B^0$    & $[K^0 p] \bar p$ & Need to flavor tag \\
         & $[K^+ n] \bar p$ & Only missing neutron \\
$\bar B^0$ & $[\bar K^0 \bar p] p$ & Need to flavor tag \\
           & $[K^- \bar n] p$ & See $\bar n$ annihilate \\ \hline \hline
\end{tabular}
\end{center}
\end{table}

\begin{figure}
\begin{center}
\includegraphics[height=3.7in]{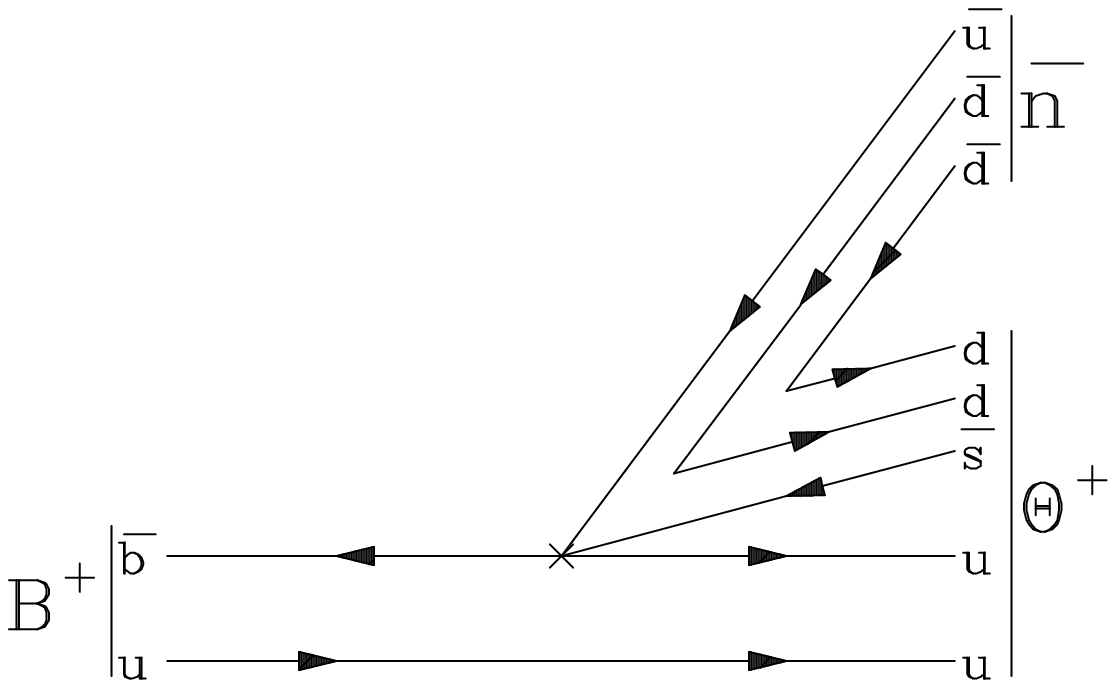}
\caption{Production of an exotic baryon $\Theta^+ = uudd \bar s$ in
$B^+$ decays. 
\label{fig:thbp}}
\end{center}
\end{figure}

\begin{figure}
\begin{center}
\includegraphics[height=3.7in]{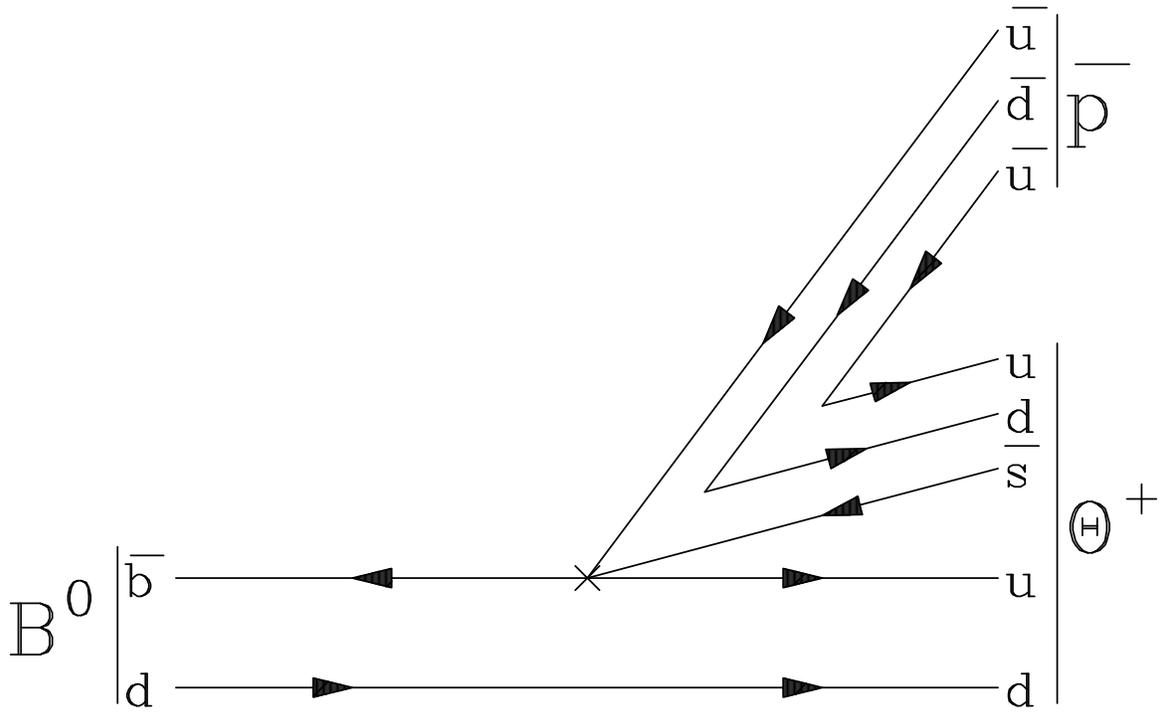}
\caption{Production of an exotic baryon $\Theta^+ = uudd \bar s$ in
$B^0$ decays.
\label{fig:thbz}}
\end{center}
\end{figure}

We summarize some decay modes of $B$ mesons useful for searching for the
$\Theta^+$ in Table \ref{tab:modes}.  Examples of quark diagrams associated
with the production of $\Theta^+$ in $B$ decays are given in Figs.\
\ref{fig:thbp} and \ref{fig:thbz}.  The weak subprocess of interest is $\bar b
\to \bar s q \bar q$, where $q=u,d$.

These figures are clearly very similar to those associated with the
production of the charmed exotic baryon depicted in the previous Section.
There may be some differences associated with differences in the structure
of the corresponding weak vertices.  In particular, in the penguin diagrams
it is not clear how directly the light $q \bar q$ pair can be associated
with the weak vertex; it may arise as a result of rescattering from the
$\bar b \to \bar s c \bar c$ subprocess, in which case Figs.\ \ref{fig:thbp}
and \ref{fig:thbz} may represent an oversimplification.
\bigskip

\centerline{\bf VI.  SPECIAL FEATURES OF {\boldmath $\Upsilon(1S)$} DECAYS}
\bigskip

The decays of $\Upsilon(1S)$ lead to large numbers of baryons in the final
state \cite{Scheck:1989aa}, and even (anti)deuterons have been observed
\cite{Albrecht:1989ag}.  Thus, it is reasonable to expect $\Upsilon(1S)$ decay
products to include multi-quark states beyond $q \bar q$ and $qqq$ if they
exist.

The neutral light-quark quantum numbers (isospin and strangeness) of the
$\Upsilon(1S)$ are an asset in searching for exotic states via missing-mass
techniques.  One simply looks for a decay $\Upsilon(1S) \to X + Y$,
reconstructing all particles in $Y$ and plotting the missing mass in $X$.
If desired, $X$ or $Y$ may be selected for the presence of a baryon or
a baryon-antibaryon pair.  Or, one may simply plot a mass of the appropriate
subsystem.

Suppose one is looking for an exotic meson with quark content $X = u u \bar d
\bar d$.  One can reconstruct the subsystem $Y$ in any of a number of ways,
such as $\rho^- \rho^-$ or $p \bar p \pi^- \pi^-$, and plot the missing mass
in $X$, or one can directly plot $M(\rho^+ \rho^+)$ or $M(p \bar p \pi^+
\pi^+)$.  In searching for $X_s = u u \bar s \bar d$ one can choose $Y$ to be
(e.g.) $\Lambda \bar p \pi^-$ or $K^- \pi^- (m \pi^0)~(m=0,1,2,\ldots)$.
One can select for a system $X$ with strangeness $S=1$, baryon number $B=1$,
and charge $Q=1$ (the quantum numbers of the reputed $\Theta^+$) by choosing
the subsystem $Y$ to have $S=-1$, $B=-1$, and $Q=-1$, for instance by taking
$Y = \bar p K^- \pi^+$.

The CLEO Collaboration has recently amassed a sample of 21 million
$\Upsilon(1S)$ decays, far exceeding that available previously
\cite{Dytman:2003qx}.  With the enhanced particle identification capabilities
and excellent neutral particle detection of the CLEO-III Detector, this
sample should be an excellent one in which to search for exotic systems with
internal quantum numbers beyond those of $q \bar q$ and $qqq$.
\bigskip

\centerline{\bf VII.  POSSIBILITIES WITH CHARM}
\bigskip

The Cabibbo-favored decays of a charmed quark, $c \to s u \bar d$, lead to
a state with three quarks of distinct flavor.  The decay $D^+ = c \bar d
\to s u \bar d \bar d$ leads to an exotic state.  The weakness of final-state
interactions in such states may be one reason why the lifetime of the
$D^+$ is longer than that of the $D^0$ or $D_s^+$, for which the
Cabibbo-favored final states are not exotic \cite{Gaillard:1974mw}.

One can then look directly for exotic final states in the decays $D^+ \to X^0 +
Y^+_{S=-1}$, where $X^0 = \gamma,~\pi^0,\ldots$ is used to measure the missing
mass of the strange subsystem $Y^+_{S=-1}$.  One has to ensure that $Y^+$ has
negative strangeness since if it is non-strange, it will not be exotic.
Similarly, the decay $D^0 \to \pi^- Y^+_{S=-1}$ can be used to probe the
missing mass of the subsystem $Y$.  Although $D$ decays do not involve enough
energy to produce exotic resonances decaying to baryon-antibaryon pairs, they
should be able to populate the lighter mass ranges proposed in Ref.\
\cite{Jaffe:1976ig}.

Decay of the $J/\psi$ into light-quark states may allow probes of the
lower mass range for exotic systems, using techniques mentioned in the
previous Section.  Even the decay $J/\psi \to \Theta^+ \bar \Theta^-$ would
be just barely kinematically allowed for a $\Theta$ of mass 1540 MeV/$c^2$.
A search for the $\Theta$ in $J/\psi$ and $\psi(2S)$ decays has revealed no
evidence for the state so far \cite{Bai:2004gk}.
\bigskip

\centerline{\bf VIII.  SUMMARY}
\bigskip

Systems containing heavy quarks are potentially a rich source of information
on exotic mesons and baryons containing lighter quarks.  We have shown how
such information may be obtained from $B$, $\Upsilon(1S)$, and charm decays.
Both $B$ decays and $\Upsilon(1S)$ provide unique advantages in searching for
the $\Theta^+(1540)$ recently reported in several experiments, since the
final states of interest -- $K^0 p$ or $K^+ n$ -- can be easily identified
under some cirumstances.  For example, $B$ decays favor a particular
strangeness ($b$ quarks give rise to $s$ quarks, while $\bar b$ quarks give
rise to $\bar s$ quarks).  The nearly hermetic detectors constructed to study
$B$ physics have some sensitivity to antineutrons, allowing the study of
the $K^- \bar n$ final state.  (Note added:  The reader's attention should
also be drawn to the recent Refs. \cite{Armstrong:2003zc,Browder:2004mp,HJLPC}
which cite the present work and offer further suggestions.)
\newpage

\centerline{\bf ACKNOWLEDGMENTS}
\bigskip

I thank Roy Briere, Sasha Glazov, Bob Jaffe, Marek Karliner, Igor Klebanov,
Peter Lepage, Harry Lipkin, Jim Napolitano, Shmuel Nussinov, Shoichi Sasaki,
Gregg Thayer, Jana Thayer, Bruce Yabsley, and John Yelton for discussions,
and the Laboratory of Elementary Particle Physics at Cornell for hospitality
during this research.  This work was supported in part by the United States
Department of Energy through Grant No.\ DE FG02 90ER40560.

\def \ajp#1#2#3{Am.\ J. Phys.\ {\bf#1}, #2 (#3)}
\def \apny#1#2#3{Ann.\ Phys.\ (N.Y.) {\bf#1}, #2 (#3)}
\def \app#1#2#3{Acta Phys.\ Polonica {\bf#1}, #2 (#3)}
\def \arnps#1#2#3{Ann.\ Rev.\ Nucl.\ Part.\ Sci.\ {\bf#1}, #2 (#3)}
\def \art{and references therein}
\def \cmts#1#2#3{Comments on Nucl.\ Part.\ Phys.\ {\bf#1}, #2 (#3)}
\def \cn{Collaboration}
\def \cp89{{\it CP Violation,} edited by C. Jarlskog (World Scientific,
Singapore, 1989)}
\def \efi{Enrico Fermi Institute Report No.\ }
\def \epjc#1#2#3{Eur.\ Phys.\ J. C {\bf#1}, #2 (#3)}
\def \f79{{\it Proceedings of the 1979 International Symposium on Lepton and
Photon Interactions at High Energies,} Fermilab, August 23-29, 1979, ed. by
T. B. W. Kirk and H. D. I. Abarbanel (Fermi National Accelerator Laboratory,
Batavia, IL, 1979}
\def \hb87{{\it Proceeding of the 1987 International Symposium on Lepton and
Photon Interactions at High Energies,} Hamburg, 1987, ed. by W. Bartel
and R. R\"uckl (Nucl.\ Phys.\ B, Proc.\ Suppl., vol.\ 3) (North-Holland,
Amsterdam, 1988)}
\def \ib{{\it ibid.}~}
\def \ibj#1#2#3{~{\bf#1}, #2 (#3)}
\def \ichep72{{\it Proceedings of the XVI International Conference on High
Energy Physics}, Chicago and Batavia, Illinois, Sept. 6 -- 13, 1972,
edited by J. D. Jackson, A. Roberts, and R. Donaldson (Fermilab, Batavia,
IL, 1972)}
\def \ijmpa#1#2#3{Int.\ J.\ Mod.\ Phys.\ A {\bf#1}, #2 (#3)}
\def \ite{{\it et al.}}
\def \jhep#1#2#3{JHEP {\bf#1}, #2 (#3)}
\def \jpb#1#2#3{J.\ Phys.\ B {\bf#1}, #2 (#3)}
\def \lg{{\it Proceedings of the XIXth International Symposium on
Lepton and Photon Interactions,} Stanford, California, August 9--14 1999,
edited by J. Jaros and M. Peskin (World Scientific, Singapore, 2000)}
\def \lkl87{{\it Selected Topics in Electroweak Interactions} (Proceedings of
the Second Lake Louise Institute on New Frontiers in Particle Physics, 15 --
21 February, 1987), edited by J. M. Cameron \ite~(World Scientific, Singapore,
1987)}
\def \kdvs#1#2#3{{Kong.\ Danske Vid.\ Selsk., Matt-fys.\ Medd.} {\bf #1},
No.\ #2 (#3)}
\def \ky85{{\it Proceedings of the International Symposium on Lepton and
Photon Interactions at High Energy,} Kyoto, Aug.~19-24, 1985, edited by M.
Konuma and K. Takahashi (Kyoto Univ., Kyoto, 1985)}
\def \mpla#1#2#3{Mod.\ Phys.\ Lett.\ A {\bf#1}, #2 (#3)}
\def \nat#1#2#3{Nature {\bf#1}, #2 (#3)}
\def \nc#1#2#3{Nuovo Cim.\ {\bf#1}, #2 (#3)}
\def \nima#1#2#3{Nucl.\ Instr.\ Meth. A {\bf#1}, #2 (#3)}
\def \np#1#2#3{Nucl.\ Phys.\ {\bf#1}, #2 (#3)}
\def \npbps#1#2#3{Nucl.\ Phys.\ B Proc.\ Suppl.\ {\bf#1}, #2 (#3)}
\def \os{XXX International Conference on High Energy Physics, Osaka, Japan,
July 27 -- August 2, 2000}
\def \PDG{Particle Data Group, K. Hagiwara \ite, \prd{66}{010001}{2002}}
\def \pisma#1#2#3#4{Pis'ma Zh.\ Eksp.\ Teor.\ Fiz.\ {\bf#1}, #2 (#3) [JETP
Lett.\ {\bf#1}, #4 (#3)]}
\def \pl#1#2#3{Phys.\ Lett.\ {\bf#1}, #2 (#3)}
\def \pla#1#2#3{Phys.\ Lett.\ A {\bf#1}, #2 (#3)}
\def \plb#1#2#3{Phys.\ Lett.\ B {\bf#1}, #2 (#3)}
\def \pr#1#2#3{Phys.\ Rev.\ {\bf#1}, #2 (#3)}
\def \prc#1#2#3{Phys.\ Rev.\ C {\bf#1}, #2 (#3)}
\def \prd#1#2#3{Phys.\ Rev.\ D {\bf#1}, #2 (#3)}
\def \prl#1#2#3{Phys.\ Rev.\ Lett.\ {\bf#1}, #2 (#3)}
\def \prp#1#2#3{Phys.\ Rep.\ {\bf#1}, #2 (#3)}
\def \ptp#1#2#3{Prog.\ Theor.\ Phys.\ {\bf#1}, #2 (#3)}
\def \rmp#1#2#3{Rev.\ Mod.\ Phys.\ {\bf#1}, #2 (#3)}
\def \rp#1{~~~~~\ldots\ldots{\rm rp~}{#1}~~~~~}
\def \rpp#1#2#3{Rep.\ Prog.\ Phys.\ {\bf#1}, #2 (#3)}
\def \sing{{\it Proceedings of the 25th International Conference on High Energy
Physics, Singapore, Aug. 2--8, 1990}, edited by. K. K. Phua and Y. Yamaguchi
(Southeast Asia Physics Association, 1991)}
\def \slc87{{\it Proceedings of the Salt Lake City Meeting} (Division of
Particles and Fields, American Physical Society, Salt Lake City, Utah, 1987),
ed. by C. DeTar and J. S. Ball (World Scientific, Singapore, 1987)}
\def \slac89{{\it Proceedings of the XIVth International Symposium on
Lepton and Photon Interactions,} Stanford, California, 1989, edited by M.
Riordan (World Scientific, Singapore, 1990)}
\def \smass82{{\it Proceedings of the 1982 DPF Summer Study on Elementary
Particle Physics and Future Facilities}, Snowmass, Colorado, edited by R.
Donaldson, R. Gustafson, and F. Paige (World Scientific, Singapore, 1982)}
\def \smass90{{\it Research Directions for the Decade} (Proceedings of the
1990 Summer Study on High Energy Physics, June 25--July 13, Snowmass, Colorado),
edited by E. L. Berger (World Scientific, Singapore, 1992)}
\def \tasi{{\it Testing the Standard Model} (Proceedings of the 1990
Theoretical Advanced Study Institute in Elementary Particle Physics, Boulder,
Colorado, 3--27 June, 1990), edited by M. Cveti\v{c} and P. Langacker
(World Scientific, Singapore, 1991)}
\def \yaf#1#2#3#4{Yad.\ Fiz.\ {\bf#1}, #2 (#3) [Sov.\ J.\ Nucl.\ Phys.\
{\bf #1}, #4 (#3)]}
\def \zhetf#1#2#3#4#5#6{Zh.\ Eksp.\ Teor.\ Fiz.\ {\bf #1}, #2 (#3) [Sov.\
Phys.\ - JETP {\bf #4}, #5 (#6)]}
\def \zpc#1#2#3{Zeit.\ Phys.\ C {\bf#1}, #2 (#3)}
\def \zpd#1#2#3{Zeit.\ Phys.\ D {\bf#1}, #2 (#3)}

\end{document}